# Quantum oscillations in 2D insulators induced by graphite gates

Jiacheng Zhu[1], Tingxin Li[1], Andrea F. Young[2], Jie Shan[1,3,4], Kin Fai Mak[1,3,4*]

[1] School of Applied and Engineering Physics, Cornell University, Ithaca, NY, USA
[2] Department of Physics, University of California, Santa Barbara, CA, USA
[3] Laboratory of Atomic and Solid State Physics, Cornell University, Ithaca, NY, USA
[4] Kavli Institute at Cornell for Nanoscale Science, Ithaca, NY, USA
[*]Email: kinfai.mak@cornell.edu

**Abstract**
We demonstrate a mechanism for magnetoresistance oscillations in insulating states of two-dimensional (2D) materials arising from the interaction of the 2D layer and proximal graphite gates. We study a series of devices based on different 2D systems, including mono- and bilayer $T_d$-$WTe_2$, angle-aligned $MoTe_2$/$WSe_2$ heterobilayers and Bernal-stacked bilayer graphene, which all share a similar graphite-gated geometry. We find that the resistivity of the 2D system generically shows quantum oscillations as a function of magnetic field corresponding to a high-density Fermi surface when they are tuned near an insulating state, in contravention of naïve band theory. Simultaneous measurement of the resistivity of the graphite gates show that these oscillations are precisely correlated with quantum oscillations in the resistivity of the graphite gates themselves. Further supporting this connection, the oscillations are quenched when the graphite gate is replaced by $TaSe_2$, a high-density metal that does not show quantum oscillations. The observed phenomenon arises from the oscillatory behavior of graphite density of states, which modulates the device capacitance and, as a consequence, the carrier density in the sample layer even when a constant electrochemical potential is maintained between the sample and the gate electrode. Oscillations are most pronounced near insulating states where the resistivity is strongly density dependent. Our study suggests a unified mechanism for quantum oscillations in graphite-gated 2D insulators based on sample-gate coupling.

The experimental observation of quantum oscillations in insulators has challenged the band theory of solids [1, 2]. Proposals involving neutral fermions [3, 4], excitons [5, 6, 7] and inverted bands [8, 9, 10, 11] have been put forth to explain the puzzling observations. Recently, Wang et al. [12] reported magnetoresistance (MR) oscillations near the insulating state of monolayer $WTe_2$ encapsulated between thin hexagonal boron nitride (hBN) dielectrics (< 10 nm) and top and bottom graphite gates. They interpreted the results as evidence for charge-neutral fermions, engendering further theoretical proposals to explain the results [6, 7, 13]. However, the graphite gates play a crucial role for the observed MR oscillations as the replacement of the graphite gates by metallic gates causes a drastic decline in the quality of the oscillations (Extended Data Fig. 10 in Ref. [12]). Moreover, some devices show two distinct oscillation frequencies with each frequency solely controlled by one of the two graphite gates (Extended Data Fig. 4 in Ref. [12]). These results call for closer scrutiny of the role of the graphite, which is itself a high-quality two dimensional electronic system.

In this letter we demonstrate a capacitive mechanism by which MR oscillations in the sample are generated by oscillations in the graphite density of states (DOS). The basic idea is illustrated in Fig. 1a, b. The two-dimensional (2D) sample and the graphite gate form the two plates of a capacitor. The total capacitance of the device is determined by the geometrical capacitance and the quantum capacitances of the 2D sample and the graphite gate. An external gate voltage maintains a constant electrochemical potential difference between the two plates. Because the quantum capacitance of the graphite gate oscillates under a perpendicular magnetic field (due to the formation of Landau levels), the total capacitance and therefore the carrier density in the 2D sample oscillate accordingly; the sample and the gate cannot be treated as independent of each other [14, 15]. Large MR oscillations near an insulating state of the sample, where the resistance is strongly dependent on the carrier density, are therefore expected. We demonstrate in this study the ubiquity of this phenomenon in various 2D materials, including mono- and bilayer $WTe_2$, angle-aligned $MoTe_2/WSe_2$ heterobilayers and Bernal-stacked bilayer graphene. Coincident MR oscillations with identical gate voltage dependence of the oscillation frequency between the sample and the graphite gate are observed. The oscillations are quenched when the graphite gate is replaced by a $TaSe_2$ gate, a 2D metal with negligible quantum capacitance oscillations in magnetic fields examined in this study. Furthermore, a 180-degree phase shift in the MR oscillations between slightly electron- and slightly hole-doped bilayer graphene is observed. The results are fully consistent with the physical picture presented in Fig. 1a, b.

All devices examined in this study are dual-gated devices with the 2D sample of interest encapsulated between hBN dielectrics and top and bottom gates. The 2D sample is contacted by conducting electrodes of the appropriate metals, which will be detailed below, in order to minimize the contact resistance. We mainly focus on few-layer graphite as the gate material and use $TaSe_2$ only in one occasion as a control experiment. The two gates allow independent tuning of the carrier density in the 2D sample and the carrier density difference between the gates (i.e. the perpendicular electric field). We also intentionally choose thinner hBN dielectric (~ 5 nm) for the top gate in order to isolate the effects from the top gate by increasing its capacitive coupling to the sample. Details of device fabrications are generic and have been reported elsewhere [16, 17, 18].

We first examine monolayer WTe$_2$ devices. The formation of electrical contacts to the bulk of monolayer WTe$_2$ requires extra care because the material is a quantum spin Hall insulator with helical edge states that can short the bulk electrical conduction at low temperatures [19, 20, 21]. We overcome this issue by inserting a thin hBN barrier in between the electrode and the monolayer WTe$_2$ in one of the contacts so that the electrode is electrically isolated from the sample edge and bulk conduction can be accessed (Fig. 1a). We use few-layer semimetallic WTe$_2$ as the contact electrode to reduce the contact resistance. We also insert few-layer graphite probes to contact the graphite top gate in order to simultaneously measure its resistance.

Figure 1c shows the top gate voltage ($V_{tg}$) dependence of the two-point resistance of monolayer WTe$_2$ at zero magnetic field and varying temperatures. The bottom gate voltage is fixed at $V_{bg}$ = -3 V. A resistance plateau (~ 200 k$\Omega$) nearly independent of temperature below 10 K is observed near $V_{tg}$ = 0 V, where WTe$_2$ is close to the charge neutrality point. This is in contrast to the insulating behavior slightly away from charge neutrality, where the resistance increases (beyond 200 k$\Omega$) with decreasing temperature (the insulating-like behavior at high electron and hole doping is caused by the increase in contact resistance at low temperatures in a two-point measurement). The result can be explained by the presence of lurking helical edge states that electrically short the insulating bulk of charge neutral WTe$_2$, as have been reported by a recent study [22]. The presence of lurking edge states in the sample is supported by the same plot as Fig. 1c under a finite perpendicular magnetic field B = 5 T (Fig. 1d). The resistance no longer saturates at low temperatures; it displays the expected insulating behavior and reaches ~ 100 M$\Omega$ at 1.8 K. Because the external magnetic field breaks time reversal symmetry and induces significant back scattering in the helical edge states, lurking edge state conduction is no longer important and bulk conduction dominates.

Figure 1e shows the MR oscillations at 1.8 K for both the top graphite gate and slightly hole-doped WTe$_2$ at $V_{tg}$ = -1.3 V and $V_{bg}$ = -3.0 V, where WTe$_2$ displays insulating behavior. A smooth magnetic field dependent resistance background is subtracted from the raw data to highlight the MR oscillations (Fig. S1). Unless otherwise specified, we will present the background-subtracted data at 1.8 K from now on. The observed MR oscillations in WTe$_2$ are consistent with the finding of Wang et al. [12]. The oscillations are almost perfectly correlated with those in the top graphite gate over a wide range of magnetic fields; the resistance dips in WTe$_2$ align with the resistance peaks in the graphite gate over multiple occurrences as marked by the dashed lines.

To further illustrate the nearly perfect correlation in the observed MR oscillations, we examine the gate voltage dependence of the MR oscillations from both monolayer WTe$_2$ and the top graphite gate in Fig. 2. Figure 2a and 2b show in contour plots the $V_{tg}$ dependence (at fixed $V_{bg}$ = -3.0 V) of the MR oscillations for slightly hole-doped WTe$_2$ and the top graphite gate, respectively. (See Fig. S2 for MR oscillations on the electron-doping side.) The graphite oscillations show a Landau fan with levels converging to near $V_{tg} \approx 0$ V at $B = 0$ T, as expected for typical quantum oscillations in graphite [23]. An almost identical Landau fan structure with levels converging to the same $V_{tg}$ and $B$ is also observed in $WTe_2$. (We cannot observe reliable MR oscillations for $V_{tg} > -1.5$ V because of the large resistance fluctuations near charge neutrality in $WTe_2$; the fluctuations are likely caused by the lurking edge states.) The nearly identical Landau fan structure of the two is further confirmed by the fast Fourier transform (FFT) of the data in Fig. 2a and 2b (Fig. S3a and S3b). The oscillation frequency displays nearly identical $V_{tg}$ dependence for the two cases.

We also examine the dependence of the MR oscillations on the bottom gate voltage $V_{bg}$ (at fixed $V_{tg} = -1.8$ V) in Fig. 2c and 2d. The MR oscillations for both monolayer $WTe_2$ and the top graphite gate are independent of $V_{bg}$, which can also be seen in the FFT data in Fig. S3c and S3d. The absence of $V_{bg}$ dependence in $WTe_2$ even though $V_{bg}$ modulates its hole density shows that the MR oscillations are not intrinsic to $WTe_2$ but are induced by the top graphite gate. The $WTe_2$ monolayer largely screens out the electric field from the bottom gate to the top gate so that the carrier density and the MR oscillations of the top gate are independent of $V_{bg}$. The result is consistent with our intentional choice of thinner top gate dielectric for stronger sample-gate capacitive coupling in order to isolate effects only from the top gate.

Next we demonstrate the ubiquity of MR oscillations near the insulating states in several 2D materials induced by capacitively coupled graphite gates. We first examine bilayer $WTe_2$ contacted by semimetallic few-layer $WTe_2$. Whereas monolayer $WTe_2$ is a quantum spin Hall insulator, bilayer $WTe_2$ is a topologically trivial insulator without helical edge states [19, 20]. The inset of Fig. 3a shows the doping dependent two-point resistance at zero magnetic fields. It shows a resistance peak exceeding 10 MΩ at charge neutrality, clearly demonstrating its insulating character. Near charge neutrality (the arrow in the inset), clear MR oscillations (without background subtraction) similar to those in monolayer $WTe_2$ are observed (Fig. 3a).

We also examined graphite-gated angle-aligned $MoTe_2/WSe_2$ heterobilayers, which form moiré superlattices because of the finite lattice mismatch between the two materials [18, 24, 25, 26]. The physics of the system can be largely captured by a single-band Hubbard model [18, 27, 28]. The inset of Fig. 3b shows the doping dependent four-point resistance at zero magnetic field. Pt contacts have been employed to reduce the contact resistance [18]. Two prominent resistance peaks at filling factor 1 and 2 are observed. They correspond to the Mott and the band insulating state, respectively [18]. A recent study has shown MR oscillations near the Mott insulating state [18]. In Fig. 3b we observe similar MR oscillations (without background subtraction) near the band insulating state (the arrow in the inset). To further illustrate the necessity of the graphite gate in inducing the observed MR oscillations, we replace the top graphite gate by a metallic few-layer $TaSe_2$ gate in a different device. Similar doping-dependent resistance compared to the graphite-gated device is seen in the inset of Fig. 3c. However, no MR oscillations can be observed near the band insulating state (Fig. 3c). Because of the much larger band mass and the lower electron mobility compared to graphite, $TaSe_2$ shows negligible DOS oscillations under magnetic fields in this study [29] and therefore cannot induce MR oscillations in the sample. The result unambiguously confirms that the observed MR oscillations are induced by the nearby top graphite gate.

The last example we will examine is Bernal-stacked bilayer graphene, which becomes a band insulator under a perpendicular electric field [30]. The high material quality allows us to study more quantitatively the underlying mechanism for the MR oscillations. Figure 4a shows a 2D map of the four-point longitudinal resistance as a function of $V_{tg}$ and $V_{bg}$. The arrows show the density and the electric field directions. The resistance maximum along the electric field direction corresponds to charge neutral bilayer graphene. The resistance at charge-neutrality increases with increasing electric field, consistent with the opening of an energy gap [31]. The vertical feature near $V_{tg} = 0$ V originates from a small region in the channel that does not overlap with the bottom gate. A representative density dependent resistance under a constant electric field ~ 0.2 V/nm is shown in Fig. 4b. Clear MR oscillations (background subtracted) are observed for both slightly electron- and slightly hole-doped bilayer graphene (Fig. 4c, corresponding doping densities are denoted by the arrows in Fig. 4b). Both the oscillation amplitude and frequency of the two are comparable because we have fine-tuned the gate voltages to maintain a constant carrier density in the top graphite gate. Interestingly, the oscillations are phase shifted by 180-degree between electron and hole doping; the dips for electron doping are aligned with the peaks for hole doping as denoted by the vertical dashed lines in Fig. 4c.

The ubiquity of the observed MR oscillations across different materials suggests an origin in the common graphite gated device architecture (Fig. 1a). Our devices are typically asymmetric, with a bottom gate dielectric that is much thicker than the top gate dielectric. We therefore ignore the bottom gate for simplicity, and consider the capacitive coupling between the sample layer and the nearby top graphite gate. Crucially, in addition to the geometric capacitance, the quantum capacitance—arising from the finite electronic compressibility—of both the sample *and the graphite gate* must be accounted for. The total capacitance between the sample and graphite gate may be written as $C = \left(C_g^{-1} + C_{Qg}^{-1} + C_{Qs}^{-1}\right)^{-1}$, where $C_g$ is the geometric capacitance ($C_{Qs}$ is the quantum capacitance of the sample, and $C_{Qg}$ is the quantum capacitance of the graphite gate). The carrier density under a constant electrochemical potential difference ($V_{tg}$ in our measurements) between the sample and the gate is $n = CV_{tg}$.

The observed oscillations can be tied to the effect of the graphite compressibility on the charge carrier density of the sample layer. The low disorder and low effective mass of graphite mean that even at low magnetic fields, the graphite compressibility oscillates. This generates magneto-oscillations in $C_{Qg}$ and consequently $C$, which in turn give rise to an oscillating sample carrier density, which we denote $\Delta n$. As a result, the sample resistance also oscillates as $\Delta R \approx \frac{dR}{dn}\Delta n$, where $\frac{dR}{dn}$ captures the density dependence of the resistivity. Notably, this effect is independent of oscillatory contributions to the sample quantum capacitance, $C_{Qs}$. It can be expected to apply to insulating regimes where sample carriers are localized and MR oscillations are not, otherwise expected.

This simple picture explains the ubiquitous MR oscillations in graphite-gated devices near the insulating states, where $\frac{dR}{dn}$ is large. It also explains the quickly diminishing MR oscillations away from the insulating states (Fig. 2 and Fig. S4), where $\frac{dR}{dn}$ substantially drops in magnitude. In other words, the sample can only "sense" the oscillations in $C_{Qg}$ when $\frac{dR}{dn}$ is large, i.e. near an insulating state. The simple picture also explains the 180-degree phase shift in the MR oscillations shown in Fig. 4c, which arise due to the sign change in $\frac{dR}{dn}$ between electron- to hole-doping ($\Delta n$ remains nearly unchanged because we have kept a constant carrier density in the top graphite gate in Fig. 4c). We can further calibrate the magnetic-field-induced density oscillations in the bilayer graphene (Fig. S5).

We conclude by summarizing the necessary conditions for observing strong MR oscillations induced by the sample-gate capacitive coupling. First, the sample-gate separation needs to be small so that the geometrical capacitance $C_g$ is large, amplifying the effects of $C_{Qg}$. Second, the sample needs to be near an insulating state where $\frac{dR}{dn}$ is large in order to amplify the MR oscillations. Finally, the gate material needs to have high enough electron mobility, and low enough effective mass, to exhibit strong DOS oscillations under moderate magnetic fields (e.g. graphite). Capacitively induced MR oscillations are expected when these conditions are met. However, we note that graphite gates may induce quantum oscillations in more subtle ways as well. For instance, compressibility oscillations in a proximal graphite gate may also modulate the screening of Coulomb repulsion in the 2D layer [24]. In small gap semiconductors, where Coulomb repulsion contributes significantly to the activation gap [32, 33], this may lead to an additional effect capable of generating MR oscillations. In this picture, as the magnetic field is tuned, the compressibility oscillations induce oscillations in the activation gaps, and consequently the resistivity of the proximal semiconductor. In light of these potential mechanisms, caution is warranted in interpreting MR oscillations in insulating samples in terms of unproven mechanisms.

**Acknowledgements**
The authors acknowledge helpful discussions with Michael Zaletel and David Cobden. J. Z. acknowledges discussions and experimental assistance with Kaifei Kang and Wenjin Zhao.

**Figures**

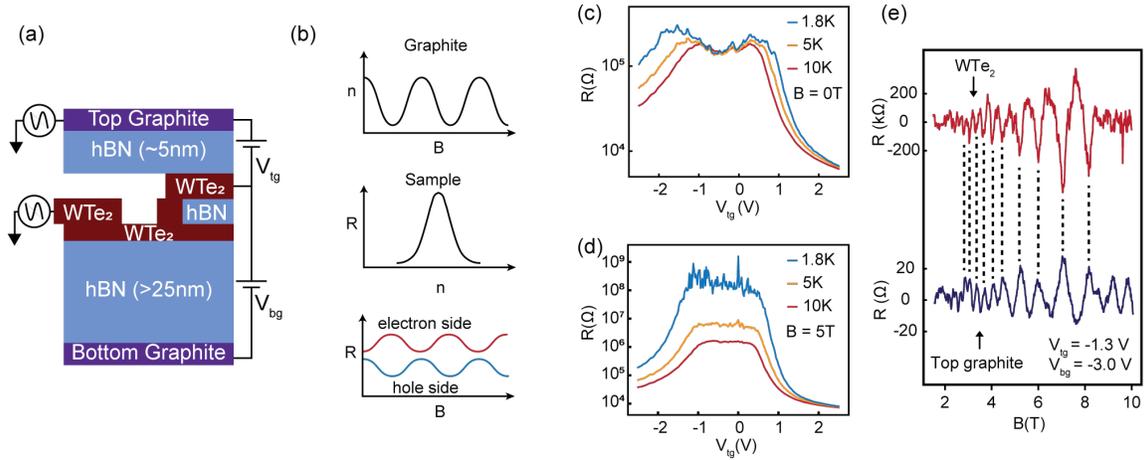

Figure 1. (a) Schematic cross-section for graphite-gated monolayer WTe$_2$ devices. The electrical connections are shown. Few-layer WTe$_2$ are used as electrical contacts. A thin hBN barrier is inserted between monolayer and few-layer WTe$_2$ to avoid direct edge contact. (b) Schematic sample-gate capacitive coupling to induce MR oscillations. Top: carrier density oscillations in graphite versus magnetic field. Middle: typical sample resistance versus carrier density near an insulating state. Bottom: MR oscillations for a slightly electron- and slightly hole-doped sample, showing 180-degree phase shift between the two. (c, d) Dependence of the two-point resistance of monolayer WTe$_2$ on $V_{tg}$ at varying temperatures under B = 0 T (c) and B = 5 T (d). The bottom gate voltage is fixed at $V_{bg}$ = -3 V. The perpendicular magnetic field suppresses conductions from the lurking helical edge states in the bulk of monolayer WTe$_2$ and increases the in-gap resistance by orders of magnitude (see text). (e) Background-subtracted MR oscillations from slightly hole-doped monolayer WTe$_2$ and the top graphite at T = 1.8 K. Dashed lines indicate the nearly perfect correlation of the two.

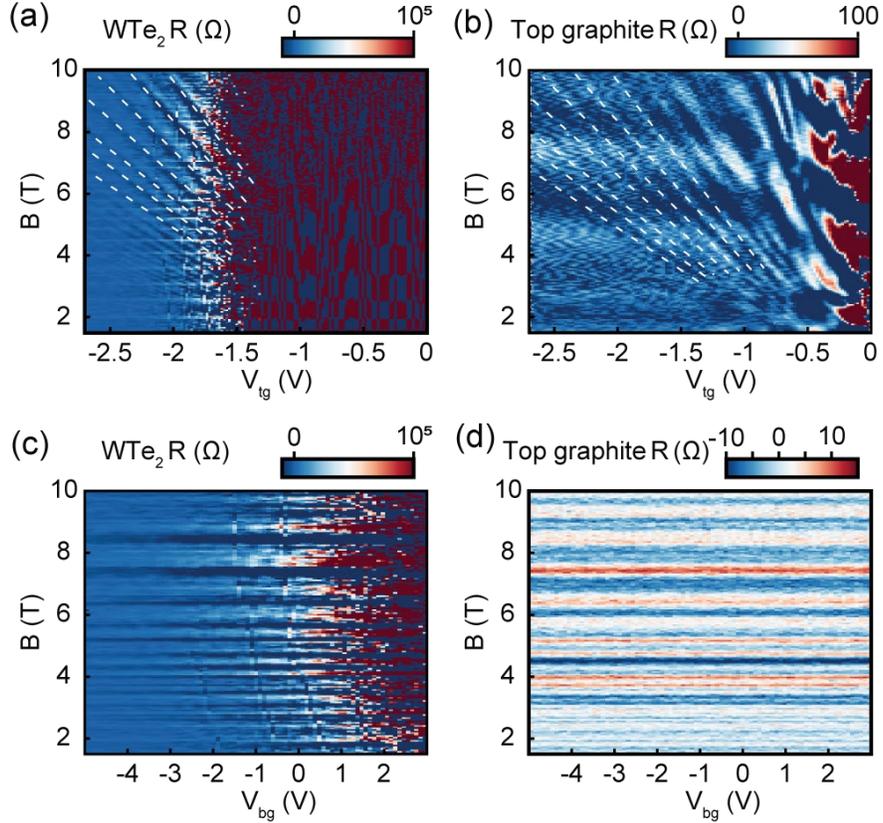

Figure 2. (a, b) Top gate voltage dependence of the MR oscillations from monolayer WTe$_2$ (a) and the top graphite gate (b) at T = 1.8 K. The bottom gate voltage is fixed at V$_{bg}$ = 0 V. Nearly identical Landau fans originating from V$_{tg}$ ≈ 0 V at B = 0 T are observed for both. MR oscillations cannot be detected in the WTe$_2$ monolayer for V$_{tg}$ > -1.5 V because of the large in-gap resistance fluctuations. White dashed lines mark a few prominent correlations of the two. (c, d) The corresponding bottom gate voltage dependence of the MR oscillations from monolayer WTe$_2$ (c) and the top graphite gate (d). The top gate voltage is fixed at V$_{tg}$ = -1.8 V. The MR oscillations are independent of V$_{bg}$ even though WTe$_2$ becomes nearly charge neutral for V$_{bg}$ > 1 V. It shows that the MR oscillations are not intrinsic to WTe$_2$.

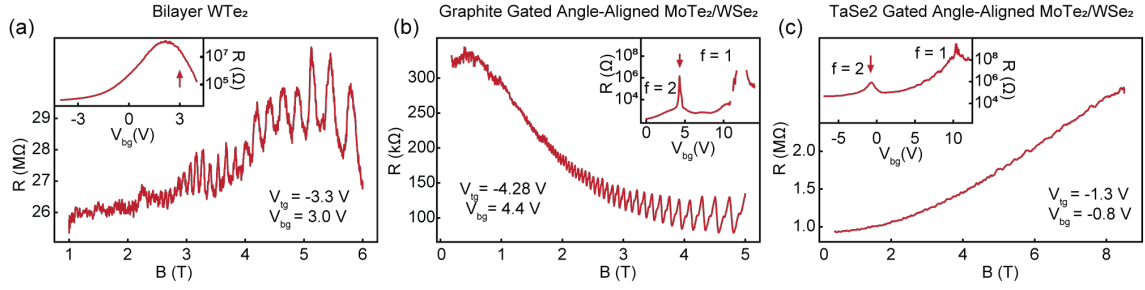

Figure 3. (a-c) MR oscillations measured at T = 1.8 K for graphite-gated bilayer WTe$_2$ (a), graphite-gated (b) and TaSe$_2$-gated (c) angle-aligned MoTe$_2$/WSe$_2$ heterobilayer. Inset of (a): Two-point resistance versus $V_{bg}$ at $V_{tg}$ = -3.3 V. Inset of (b): Four-point resistance versus $V_{bg}$ at $V_{tg}$ = -4.28 V. Inset of (c): Two-point resistance versus $V_{bg}$ at $V_{tg}$ = -1.3 V. The arrows mark where the MR oscillations are measured. The filling factor 1 and 2 insulating states are labeled in the last two insets.

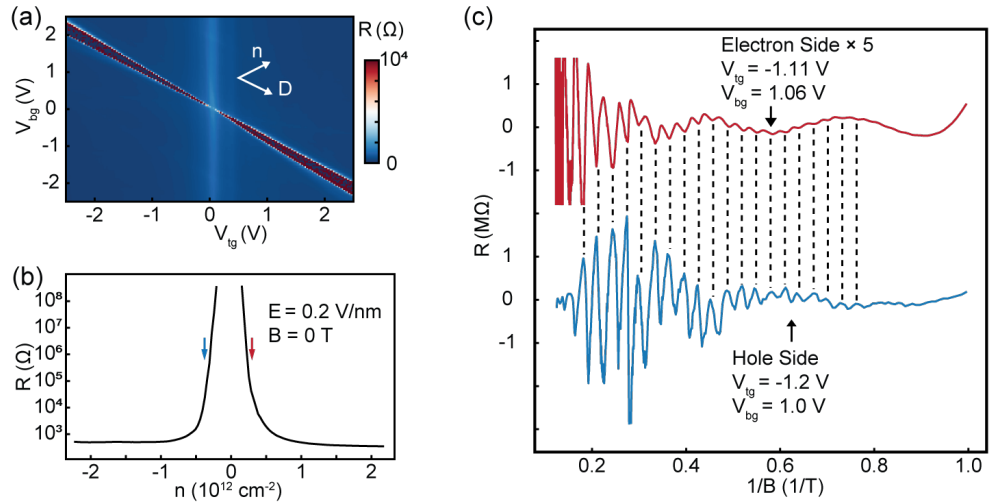

Figure 4. (a) 2D map of the four-point longitudinal resistance in Bernal-stacked bilayer graphene as a function of $V_{tg}$ and $V_{bg}$ at 1.8 K. The density and electric field axes are labeled. (b) Doping density dependent resistance under a constant electric field ~ 0.2 V/nm. The arrows mark where the MR oscillations in (c) are measured. (c) MR oscillations versus inverse magnetic field for slightly electron- and slightly hole-doped bilayer graphene. The vertical dashed lines correspond to an oscillation frequency of 31.9 T. The oscillations between electron and hole sides are 180-degree shifted.

# Supplemental materials

## Methods

Devices of van der Waals heterostructures were assembled by the layer-by-layer dry transfer method [1]. Transport measurements were performed in a Quantum Design Physical Property Measurement System (PPMS) and a He-4 Cryomagnetic cryostat. Resistance measurements were performed using standard low-frequency lock-in techniques. The excitation frequency is between 13.33 Hz and 83.33 Hz and the excitation amplitude is between 0.2 mV and 1 mV.

## Background subtraction in monolayer WTe$_2$

To extract the oscillatory part of the magnetoresistance, we fit the raw data to an $11^{th}$-order polynomial. Figure S1a and S1b show, respectively, the raw MR data of the top graphite gate and the WTe$_2$ monolayer corresponding to Fig. 1e in the main text. The $11^{th}$-order polynomial fit to the smooth background is also shown for each case.

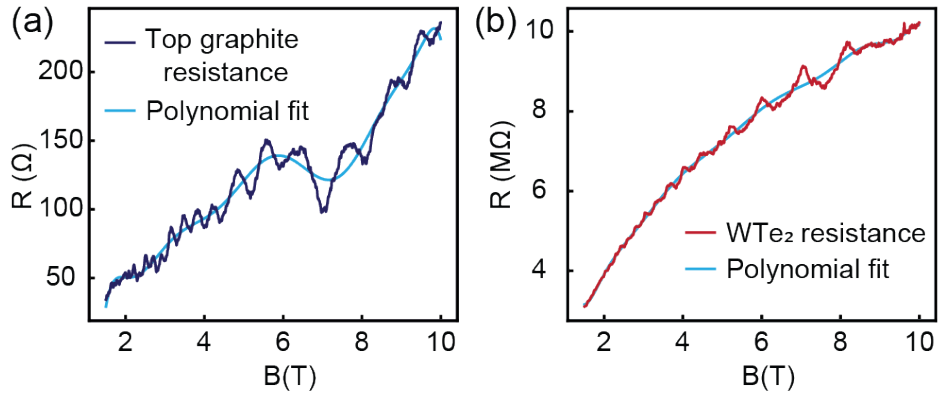

Figure S1. (a, b) Raw MR data for the top graphite gate (a) and the WTe$_2$ monolayer (b) corresponding to Fig. 1e in the main text. The cyan curves are the polynomial fits to the smooth background.

## MR oscillations in WTe$_2$ on the electron doping side

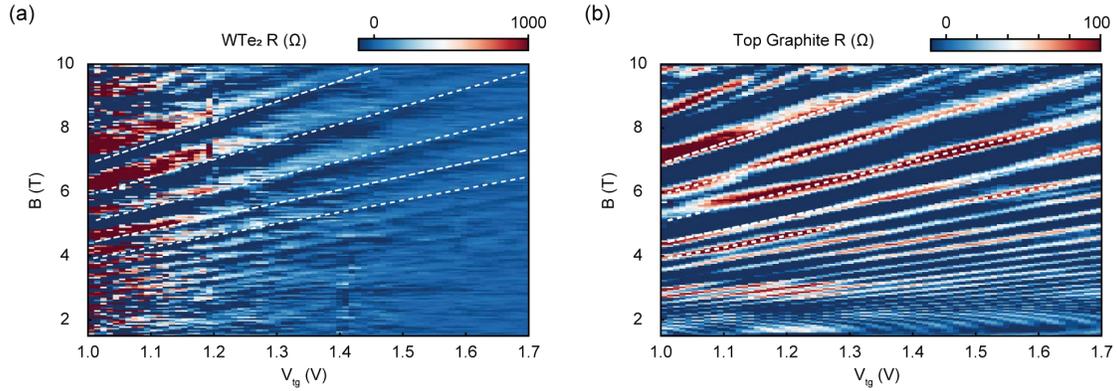

Figure S2. (a, b) Top gate voltage dependence of the MR oscillations from electron-doped monolayer WTe$_2$ (a) and the top graphite gate (b) at T = 1.8 K. The bottom gate voltage is fixed at $V_{bg}$ = 0 V. Nearly identical Landau fans originating from $V_{tg}$ = 0 V and B = 0 T are observed for both. White dashed lines mark observed correlation in MR. MR oscillations cannot be detected in the WTe$_2$ monolayer for $V_{tg}$ < 1 V because of the large in-gap resistance fluctuations. Here the phase of the oscillations for the two is the same for both electron and hole doping (see Fig. 2) because the top graphite gate changes from hole doping here to electron doping in Fig. 2.

# FFT analysis

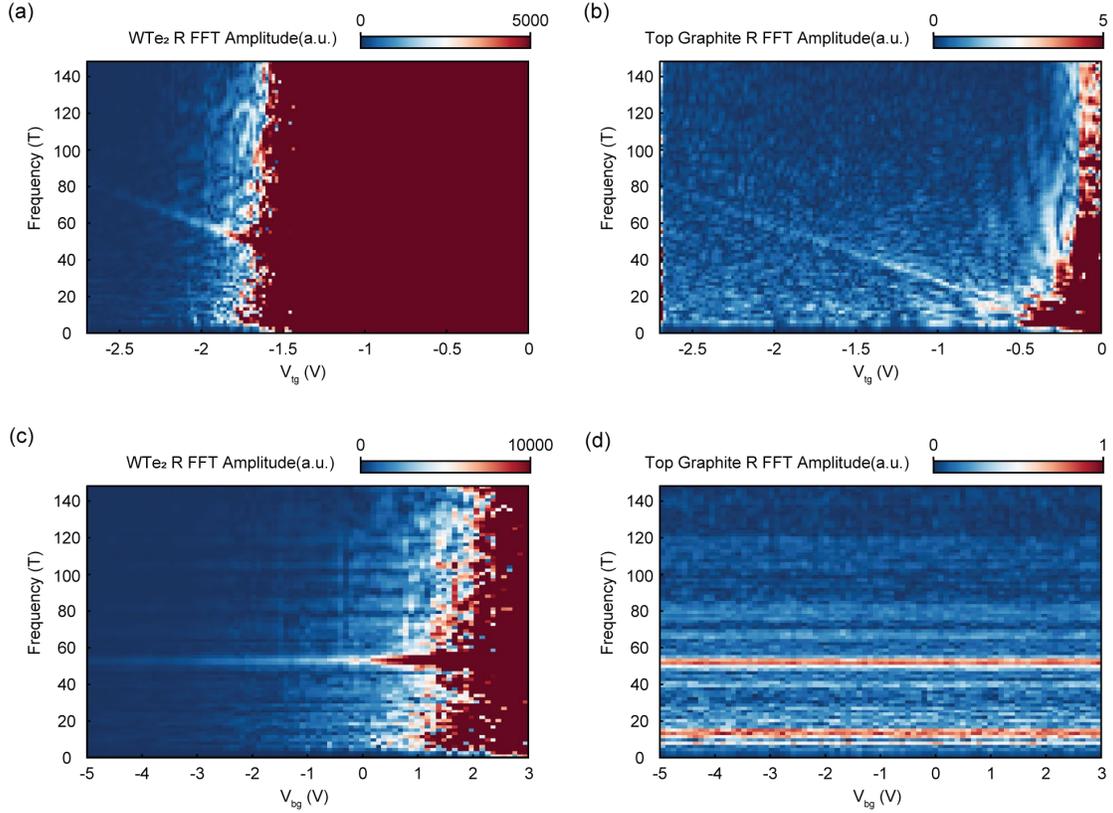

Figure S3. (a-d) FFT amplitude of the MR oscillations in monolayer WTe$_2$ and the top graphite gate corresponding to the data in Fig. 2 in the main text. The $V_{tg}$ dependence is shown in (a) and (b) for WTe$_2$ and graphite, respectively. The bottom gate is fixed at $V_{bg}$ = 0 V. A resonance linearly dependent on $V_{tg}$ emerging from $V_{tg}$ = 0 V is observed. The $V_{tg}$ dependence of the resonance for the two cases are nearly identical. The $V_{bg}$ dependence is shown in (c) and (d) for WTe$_2$ and graphite, respectively. The top gate is fixed at $V_{tg}$ = -1.8 V. The resonance is independent of $V_{bg}$ in both cases.

## Comparing $\Delta R$ and $\frac{dR}{dn}$

To further confirm our interpretation based on sample-gate capacitive coupling, we measure the magnetoresistance as a function of doping density in the bilayer graphene device near an electric field ~ 0.2 V/nm. At each doping density, we subtract a smooth background from the magnetoresistance to obtain $\Delta R$. We also evaluate $\frac{dR}{dn}$ at each magnetic field. Figure S4 shows an example of the doping dependence of $\Delta R$ and $\frac{dR}{dn}$ at a fixed magnetic field 4.4 T for both electron and hole doping. In agreement with our physical picture presented in the main text, $\Delta R$ and $\frac{dR}{dn}$ essentially show the same density dependence.

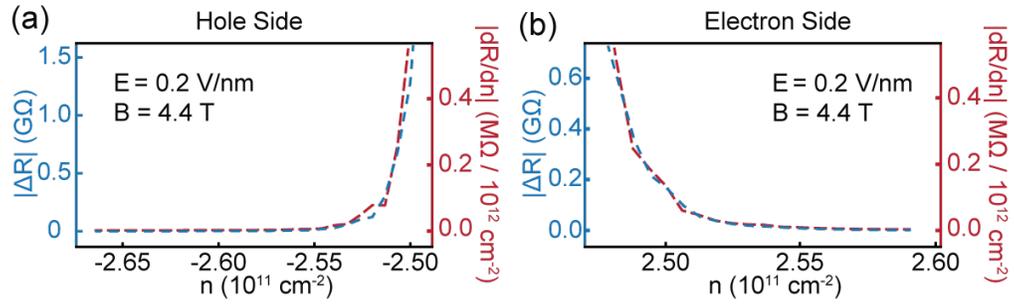

Figure S4. Doping dependence of the absolute value of $\Delta R$ and $\frac{dR}{dn}$ at a fixed magnetic field 4.4 T and near an electric field 0.2 V/nm for both hole (a) and electron (b) doping.

# Extracting density oscillations

To extract the density oscillations in the bilayer graphene induced by the top graphite gate, we measure the sample resistance versus density at each magnetic field, which allows us to extract $\frac{dR}{dn}$ versus magnetic field. The oscillation in density $\Delta n$ can then be obtained by using the relation $\Delta n = \Delta R / \frac{dR}{dn}$. Here $\Delta R$ is the background-subtracted MR oscillations at a given nominal density $n$ determined by the geometrical capacitance of the device. We have used the magnetic field dependent $\frac{dR}{dn}$ at this density $n$ to calculate $\Delta n$. An example is shown in Fig. S5.

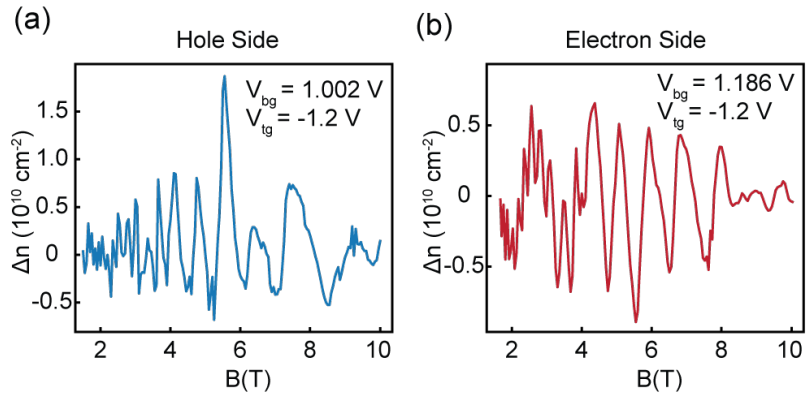

Figure S5. Extracted carrier density oscillations in slightly hole-doped (a) and slightly electron-doped (b) bilayer graphene versus magnetic field.